\documentclass[smallextended,]{article}
\usepackage{graphicx}
\usepackage{multirow} 
\usepackage{float, epsfig, latexsym, graphicx, epsf, rotate, amsfonts, setspace,morefloats,epstopdf,pdflscape,subfig}
\usepackage{tikz}
\usetikzlibrary{calc,fit,spy}
\usepackage{color}
\usepackage[font=small,labelfont=bf]{caption}

\usepackage{textgreek}
\usepackage{amsmath, amssymb, bm} 
\usepackage{gensymb}
\usepackage{parskip}
\graphicspath{{./pics/}}

\usepackage{makeidx}
\usepackage{nomencl}
\makenomenclature
\makeindex

\usepackage{lipsum}
\usepackage[lmargin=1in,rmargin=1in,tmargin=1in,bmargin=0.8in]{geometry}
\newcommand\grouphead[1]
{%
  \rlap{%
    \parbox{\textwidth}{\textbf{\large#1 \hrulefill}}%
  }\hfill
}
\renewcommand{\nomgroup}[1]
{%
  \ifx#1A\relax
    \item[\grouphead{\normalsize{General Symbols}}]
  \fi
  \ifx#1G\relax
    \item[\grouphead{\normalsize{Greek Symbols}}]
  \fi
   \ifx#1Z\relax
    \item[\grouphead{\normalsize{Superscripts }}]
  \fi
  }

\usepackage[authoryear,round]{natbib}
\usepackage{titlesec}
\titlespacing\section{0pt}{6pt plus 4pt minus 2pt}{0pt plus 2pt minus 2pt}
\titlespacing\subsection{0pt}{6pt plus 4pt minus 2pt}{0pt plus 2pt minus 2pt}

\begin{document}
\title{A data-driven proxy to Stoke's flow in porous media}
\author{Ali Takbiri-Borujeni \and Hadi Kazemi         \and   Nasser Nasrabadi   }
\maketitle

\section*{Abstract}
The objective for this work is to develop a data-driven proxy to high-fidelity numerical flow simulations using digital images. 
The proposed model can capture the flow field and permeability in a large verity of digital porous media based on solid grain geometry and pore size distribution by detailed analyses of the local pore geometry and the local flow fields. 
To develop the model, the detailed pore space geometry and simulation runs data from 3500 two-dimensional high-fidelity Lattice Boltzmann simulation runs are used to train and to predict the solutions with a high accuracy in much less computational time. 
The proposed methodology harness the enormous amount of generated data from high-fidelity flow simulations to decode the often under-utilized patterns in simulations and to accurately predict solutions to new cases.
The developed model can truly capture the physics of the problem and enhance prediction capabilities of the simulations at a much lower cost. 
These predictive models, in essence, do not spatio-temporally reduce the order of the problem. 
They, however, possess the same numerical resolutions as their Lattice Boltzmann simulations equivalents do with the great advantage that their solutions can be achieved by significant reduction in computational costs (speed and memory).
  
\noindent Source code: \url{https://github.com/Ali805509/A-data-driven-proxy-to-Stokes-flow-in-porous-media} 

\noindent  Velocity dataset:   \url{https://doi.org/10.6084/m9.figshare.7889759}

\noindent Geometry dataset:   \url{https://doi.org/10.6084/m9.figshare.7889741}


\section{Introduction}
\subsection{Imaged-based flow simulations}

Darcy's principles \citep{darcy1856fontaines} describe the fluid flow of single-phase fluids in porous media at low Reynolds numbers, which is of significant importance in earth sciences, hydrology, and petroleum engineering. According to Darcy equation, pressure gradients are linearly proportional to the fluid rate; the proportionality constant is permeability, which is merely a function of pore space topology of porous media irrespective of the fluid type. In numerical flow simulators for porous media, permeability values are obtained based on the data collected from the field and experiments. An accurate quantification of permeability is difficult due to the variations in pore space morphology characteristics. Permeability have been obtained from experiments and also from analytical and empirical expressions that relate permeability to some attributes of the porous media, such as porosity and pore size distribution. The analytical expressions are, however, only approximations for ideal cases while the empirical expressions have utility only in media similar to scenarios for which they were obtained and thus, are inaccurate when applied to a wide range of other media. Experimental approaches are generally preferred if it is not possible to account for all relevant physics by an equation or model; however, they tend to be time consuming and expensive. Furthermore, they do not capture the effect of pore space morphology characteristics on the flow field and thus, on the permeability. For certain properties, such as permeability, tortuosity, and inertial factors of the porous media, high-fidelity numerical simulations using digital images have become a credible alternative, enabled by improvements of 3D imaging techniques, numerical methods, and computing power \citep{Takbiri-Borujeni2013,sanematsu2015image}. Appealing aspects of this approach include the ability to probe pore-scale physics at a level not possible with traditional experiments and the ability to perform an endless set of numerical tests without degrading or altering the sample. There are considerations that can limit this digital approach including whether the imaging technique can resolve all relevant characteristic scales in the pore space and whether numerical algorithms can accurately model the physical processes. Higher resolution, however, mandates higher computation power. In high-fidelity numerical simulation models, the expensive computational costs, the intensive memory requirements, and the poor scaling performances have traditionally prevented their applications beyond toy or small-scale problems, even using the modern high performance computing systems. 

In image-based pore-scale modeling, the domain is discretized into nodes, voxels, or volume elements, and the resulting grid is used to numerically approximate the relevant partial differential equations for flow, namely computational fluid dynamics (CFD).
 There is a group of numerical modeling techniques that can utilize the voxel data from X-ray tomography or similar methods as the numerical grid. This gridding approach has become widely used in porous media studies in conjunction with the Lattice Boltzmann (LB) simulations, and has been proved to be highly effective for simulating fluid flow through porous media \citep{Takbiri-Borujeni2013}.

LB simulations have been applied to flow simulations of realistic porous media \citep{succi1989three,ferreol1995lattice,jin2004direct} with the advantage of being flexible in the specification of variables on the complex boundaries in terms of simple particle bounce back and reflection. 
This flexibility has opened up the potential in its use for modeling and simulating flow in complex media, such as porous rocks. 
Challenges for applying LB to real problems include finite-size effects and relaxation time dependence of no-flow boundaries. In image-based simulations, the accuracy of the calculated macroscopic properties depends on the spatial resolution of the rock image \citep{ginzbourg1994boundary,manwart2002lattice}.
 However, there is always a trade-off between image resolution and computational power. Furthermore, in all digital samples, there is a resolution threshold, below which certain flow characteristics, such as recirculation, are not resolved \citep{maier2010lattice}. 

The main advantage of pore-scale flow simulations is that explicit influence of each impacting factor can be studied by isolating the effect of other parameters. Attempts of this tabulation of all these impacts have not been manageable yet since such a multi-dimensional parametric study requires comprehensive efforts and time. 
  In this respect, this work aspires to change the status quo and make a transformative leap by combining pore-scale modeling with physics-based ML to develop proxy models, which can be used to determine the flow fields at very little additional cost. 
  It is also important to note that, using trained and validated physics-based data-driven proxy (PB-DDP) model will give us a luxury of performing pore-scale flow simulations, in which computational expenses are not restrictive. 
  Recently, there have been numerous studies on application of machine learning machine learning (ML) in CFD, most of which are limited to building interpretable reduced order models (ROMs). 
  In ROMs, where the number of variables is reduced to simplify the governing equations and the relationships between inputs and outputs, some details are inevitably overlooked. On the other hand, the widespread success of ML-based predictive modeling in other disciplines, such as autonomous cars, suggests a great opportunity to advances in the state-of-the-art by combining conventional CFD simulation techniques with predictive capabilities of PB-DDP models to truly capture the physics of the problem and enhance prediction capabilities of the simulations at a much lower cost. These predictive models, in essence, do not spatio-temporally reduce the order of the problem. They, however, possess the same numerical resolutions as their CFD equivalents do with the great advantage that their solutions can be achieved by significant reduction in computational costs (speed and memory). Essentially, the predictive models learn the nature of communications among grid cells and decode the spatial correlations between them (auto- and cross-correlations) in the entire computational domain and can accurately predict solutions to completely new sets of simulation runs, from beginning to end. 

Recently, deep convolutional networks (CNNs) with hierarchical feature learning capability have outperformed the state of the art in many computer vision tasks, including image classification \citep{simonyan2014very}, segmentation \citep{long2015fully}, and synthesis \citep{goodfellow2014generative}. Despite in classification tasks, where the network predict a single class label for an input image, in many visual tasks, the desired output could be a class label, or a continuous value, assigned to each pixel of the input image. 

\citet{ciresan2012deep} predicted the class label of each pixel by training a network in a sliding-window fashion which takes a patch around each pixel.
 This network, then, is able to localize and also is more robust to overfitting the training data, i.e., generated patches, is much larger than the number of training images. 
 However, this framework is quite slow due to separate processing of each patch, which results in a lot of redundancy on overlapping patches.  Moreover, such network should deal with the trade-off between the localization and context. Large patches need  many  pooling layers that can reduce the localization  performance, while small patches only  incorporate little context information in the final decision. More recent studies \citep{long2015fully, seyedhosseini2013image} proposed to fuse the fine to coarse features from multiple layers in different depth. This enables the network to achieve an accurate localization while having a large receptive field (context) at the same time. 
 In the work performed by \citet{ronneberger2015u}, the authors introduced U-Net which employed contracting path in its Auto-Encoder architecture to capture context and enable precise localization. 
Furthermore, training a very deep neural network is quite a challenging task. More specifically, it is hard for a deep network to find an optimal solution compared to shallower counterparts. One of the main issue in training a deep network is the vanishing  gradient problem, making it difficult to tune the  parameters of the early layers in the network \citep{glorot2010understanding}.
 In the past couple of years, multiple training strategies have been proposed to train a deep neural network effectively, including deep supervision in hidden layers \citep{lee2015deeply}, initialization scheme \citep{glorot2010understanding}, and  batch  normalization \citep{ioffe2015batch}. 
 \cite{he2016deep} introduced residual connections in which they employ additive merging of signals to improve the training speed, and gradient flow through the networks.

\subsection{Application of data-driven modeling in engineering problems}
Applications of ML have gained lots of popularity in the past few years throughout various industries. This rise in popularity is due to new technologies, such as sensors and high-performance computing services, e.g., Apache Hadoop and NoSQL that enable big-data acquisition and storage in different fields of studies, such as social networks, stock market, natural language processing, oil and gas industry, and automotive industry. 
The application of ML in CFD has gained considerable interest recently, mostly to build reduced order  models (ROMs), where the number of variables is reduced to define simpler relationships between inputs and outputs. 
However, in such applications of ML in CFD, it is inevitable to overlook some details. 
On the other hand, predictive ML techniques suggest a greater opportunity, when the conventional CFD simulation techniques are combined with predictive capabilities of PB-DDP models. 
Such approaches can truly capture the physics of the problem and enhance prediction capabilities of the simulations at a much lower cost.

Unlike automotive industry, the application of Artificial Intelligence (AI) in CFD has been limited to interpretable models from data \citep{shelton1993optimization,tracey2013application,kutz2017deep}, and predictive models are yet to be employed. 
The widespread success of predictive modeling in complex problems suggests a great opportunity to advances in the state-of-the-art by combining conventional CFD simulation techniques with ML predictive modeling to truly capture the physics of the problem and enhance prediction capabilities of the simulations at a much lower cost. 
This can be achieved by developing physically interpretable spatio-temporal simulations of complex CFD problems and introducing significant reduction in computational cost (speed and memory). 

\section{Samples}
In this work, LB simulations are performed over computer generated structures, which provide a number of advantages for testing pore-scale modeling algorithms. 
The most intuitive advantage is the ability to fully control the pore structure. Another advantage related to image-based modeling is that the geometric-based data, e.g., locations and sizes of solid grains in a random packing can be converted to voxel data at any desired image resolution without segmentation error. 
Computer-generated packings have been widely used to simulate granular materials. In some cases, unconsolidated sphere packs have been modified using procedures that mimic diagenetic processes, thus producing consolidated materials \citep{bosl1998study,zhan2010pore,kameda2004permeability}. 
We generated two-dimensional circle pack images of size 256x256 pixels, each consisting of 8 grains (circles) with 50 pixels diameter with random positions. 
A total of 3550 images are generated for LB simulation runs to determine the permeability.

A body force approach, which is an alternative to specifying pressure values at the inlet and outlet of the domain, is used. 
Periodic boundary condition is applied on all the external faces. 
The bounce-back boundary scheme is used to implement the no-flow boundary conditions at the void-solid interfaces.
Having a reasonably large pore- and grain-sizes of the porous media, calculation of the permeability are done without substantial numerical errors (finite-size errors and relaxation-time dependence of the no-flow boundaries) \citep{Takbiri-Borujeni2013,Takbiri-Borujeni2016}.

\section{Lattice Boltzmann Method}
The LB simulation method is based on kinetic theory and can be used to simulate many hydrodynamic systems \citep{sukop2006dt}. 
In this method, positions of particles are limited to nodes of a lattice with equal spacing. 
The LB equation with streaming and single velocity relaxation operator (LBGK) collision is
 \begin{equation}
 f_i({\bf x}+{\bf e}_i\Delta t, t+\Delta t) = f_i({\bf x}, t) - \frac{ \left(f_i({\bf x}, t)- f_i^{eq} ({\bf x}, t)\right)}{\tau} \,,
\end{equation}
\nomenclature{${ \bf e}_i$}{Directions {\it i} in which fluid particles can move}
\nomenclature{$f_i$}{Discrete distribution functions in {\it i} direction in velocity space}
\nomenclature{$f_i^{eq}$}{Discrete equilibrium distribution functions in {\it i} direction}
\nomenclature[G]{$\tau$}{Relaxation time}
\noindent\hspace*{-5pt}in which ${ \bf e}_i$ are directions in which fluid particles can move, $\tau$ is the relaxation time, $f_i$ are the discrete distribution functions in velocity space, and and $f_i^{eq}$ are the equilibrium distributions, 
 \begin{equation}
 f_i^{eq} ({\bf x}) = w_i \rho({ \bf x}) \left[ 1+ \frac{ ({\bf e}_i.{\bf u})}{{c_s}^2} + \frac{({ \bf e}_i.{\bf u })^2}{2{c_s}^4} - \frac{{\bf u}^2}{2 {c_s}^2}\right] \,,
 \label{eq}
\end{equation}
\nomenclature{$w_i $}{Weight factors for {\it i} direction in Equation~\ref{eq}}
\nomenclature{$c_s$}{Sound speed in the fluid}
\noindent\hspace*{-3pt}where $w_i $ are weight factors specific to different directions, $c_s = 1/\sqrt {3}$ is the speed of sound in the fluid, ${ \bf u}$ is the fluid velocity, and $\rho$ is the 
fluid density. 
In this work, the $D_2Q_{9}$ model (two dimensions and nine directions of fluid movement) is used. 
 Velocity vectors for this model are described below,
 \begin{equation}
e_i = \begin{Bmatrix} 
(0,0) & i = 0  \\ 
(\pm1,0),(0,\pm1); & i = 1,2,3,4  \\ 
(\pm1,\pm1); & i = 5,6,7,8  
\end{Bmatrix} \, .
\end{equation}

%
In LB simulations, parameterized values of the lattice constants and fluid in lattice units are used in simulation while correspondence between the real physical system being simulated and the parameterized simulation is achieved through Reynolds number (principle of dimensional similarity) \citep{chukwudozie2013pore}.

LBM simulations in this study are performed using the Parallel Lattice Boltzmann Solver (PALABOS, 2012). 
The PALABOS (Parallel Lattice Boltzmann Solver) code is used for solving the flow problems in this study \citep{latt2009palabos}. 
 
\subsection{Permeability.}
Permeability is calculated from \begin{equation}
K = -\mu \frac{\langle {\bf u} \rangle}{\nabla p} \,,
\label{perm}
\end{equation}
\nomenclature{$K$}{Permeability tensor of the porous medium}
\nomenclature{$\nabla p$}{Dynamic pressure gradient in the fluid}
\noindent\hspace*{-3pt}in which $k$ 
is the permeability tensor of the porous medium, $ \langle {\bf u} \rangle$ is average velocity of the fluid  in the domain, $ \mu$ is viscosity of the fluid, and $\nabla p$ is dynamic pressure gradient of the fluid. 
Velocity values in each node are computed in all directions using the LB simulations to determine the permeability tensor.

\section{Deep convolutional neural network}
In order to achieve an accurate and efficient model, we employ a deep convolutional neural network (DCNN) based on contracting paths and residual blocks. Since the network consists of only convolutional layers, it can take any arbitrary-sized image as input and generate an output of similar size. For down-sampling, we use convolutional layers with increased stride instead of pooling layers. After a series of successive strided convolution, the spatial size of feature maps become much smaller than that of the input image. To increase the computational capacity of the network, the generated feature maps by the last strided convolution is followed by multiple residual blocks before upscaling to the same size as the input image. 
The residual connections improve the gradients flow at the training time. 
Finally, to rescale the feature maps to the size of input image, we exploit Nearest Neighbor (NN) up-sampling followed by a convolutional layer, instead of deconvolutional layers \citep{zeiler2010deconvolutional} to prevent checker-board artifacts.
Generally, as we go deeper in a DCNN, the size of receptive field increases, which means the learned feature maps represent more abstract and global contextual features. However, the information about the exact local structure of the image may be lost. On the other hand, the feature maps in early layers, which have smaller receptive fields, preserve the local structure information. This information is critical for effective velocity field predictions. Consequently, to preserve the local structure information, high resolution features from the contracting path (down-sampling) are combined with the output of the NN up-sampling layer. Then, the subsequent convolution learns to produce a more precise output based on this information. Exploiting the learned discriminative features by the proposed DCNN, we can produce an accurate prediction of velocity fields.

\subsection{Architecture set-up}
\begin{figure}[h]
   \centering
   \includegraphics[width=\textwidth ,keepaspectratio]{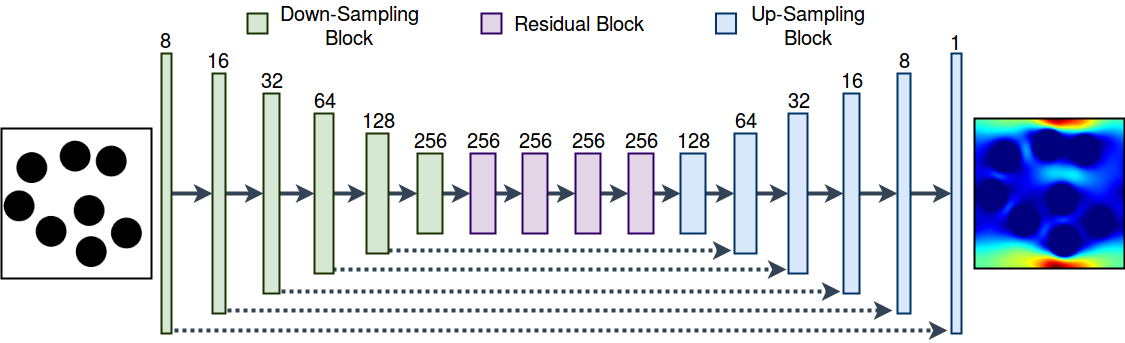}\\
   \caption{U-Net architecture.}
   \label{network}
 \end{figure}

Figure~\ref{network} shows the architecture of the proposed network. It consists of 6 strided convolutions which reduce the size of input by a factor of 64, followed by four residual blocks.  Each  residual  block  consists of  two  3x3  convolutional  layers. Finally, In  order  to  obtain  the  final  prediction map, we add six subsequent up-sampling blocks on top of the residual blocks. The input to each up-sampling block is the feature maps of the previous layer concatenated in depth with those of contracting down-sampling path. As mentioned earlier each up-sampling block comprises successive NN-upsampling and 3x3 convolution with unit stride. Note that all the convolutions are followed by a Batch Normalization and Relu activation function. 
 Since our input (geometry input image) is the simulated velocity fields we employ reflection 1x1 padding for all the convolutions.

To train the network, we first normalize the velocity maps. To  reduce  overfitting, we employed data augmentation to enlarge the training data samples. To this end, we randomly flip the maps vertically with a probability of $0.5$. The input density and their corresponding velocity maps are used to update the parameters of the network minimizing the $L_1$ norm error:
\begin{align}
	l = \sum^N_{i=1} \parallel u^i_x - \hat u^i_x(d) \parallel_1 \, ,
\end{align}
\nomenclature{$u_x$}{Calculated x-direction velocity by LB simulations}
\nomenclature{$\hat u_x$}{Estimated x-direction velocity by the network}
\nomenclature{$d$}{input map (image)}
\nomenclature{$l$}{ $L_1$ norm error}
\noindent\hspace*{-3pt}where $d$ is the input map, $\hat u_x$ is the estimated velocity map by the network,  $u_x$ is the ground truth velocity map, and $N$ is number of samples in the training dataset. 
Adam optimization technique is used with a learning rate of $1e-3$, and an $L_2$ weight decay of $2e-5$. 
The network was implemented in Pytorch running on a NVIDIA TITAN  X GPU. 
The network is trained for 100 epochs and the model with the minimum error on validation is selected.

\section{Results}
\begin{figure}[h]
   \centering
   \includegraphics[width=0.5\textwidth ,keepaspectratio]{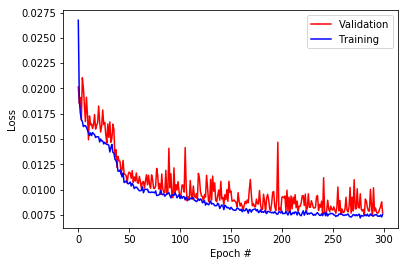}\\
   \caption{Training and validation loss curve}
   \label{loss}
 \end{figure}

To develop the model,  the velocity values for the entire images are normalized between zero and one (the minimum value the velocity values is transformed into zero, the maximum value is transformed into one, and every other value is transformed into a decimal between 0 and 1).
 All the simulation cases are divided into two sections. 
The first section with 85\% of the data (3,000 image pairs out of 3,550 total pairs) is used to train the model while the remaining 15\% of the data are used as test data. 
 The training and validation loss curve for the training process is depicted in Figure \ref{loss}.

\begin{figure}[h]
   \centering
   \includegraphics[width=\textwidth ,keepaspectratio]{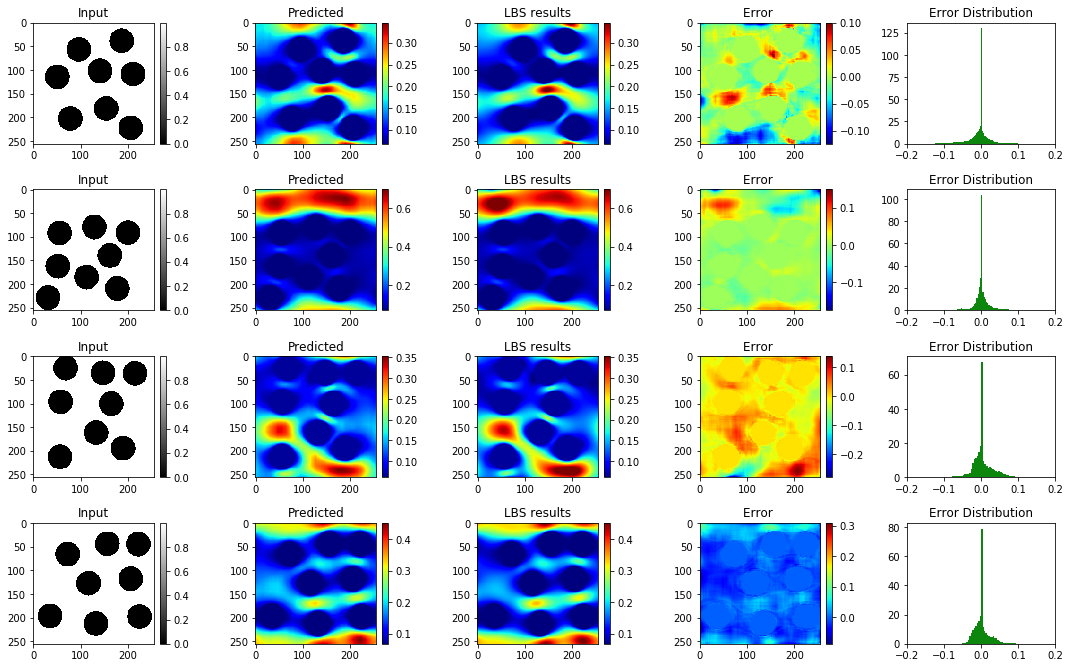}\\
   \caption{Side-by-side comparison of the normalized velocity values predicted by DDP vs. LB simulations results (ground truth) for four test cases. (a) input images used for simulations; (b) model predictions; (c) LB simulation results (ground truth); (d) error percentage between (b) and (c); and (e) distribution of error in each case.}
   \label{error}
 \end{figure}

\begin{figure}[h]
   \centering
   \includegraphics[width=\textwidth ,keepaspectratio]{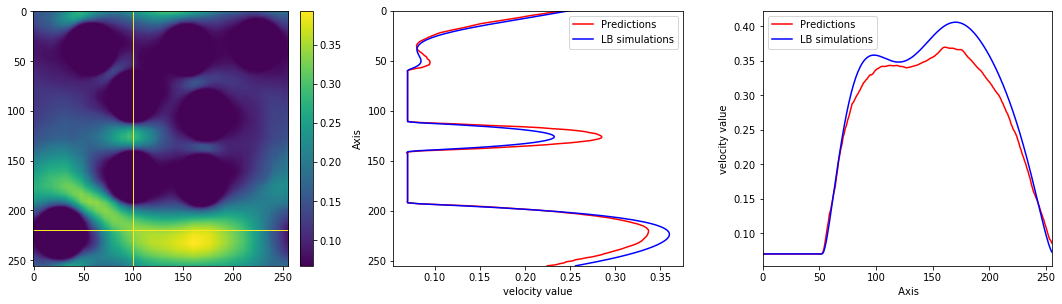}\\
   \caption{Point-by-point comparison of the normalized velocity values predicted by DDP vs. LB simulations results (ground truth) for four test cases. (a) depiction of the normalized velocity contour plots and the vertical and horizontal cross-sections; (b) velocity profiles along the vertical cross-section of the simulation domain; and (c) velocity profiles along the horizontal cross-section of the simulation domain.}
   \label{profile}
 \end{figure}

\begin{figure}[h]
   \centering
   \includegraphics[width=.5\textwidth ,keepaspectratio]{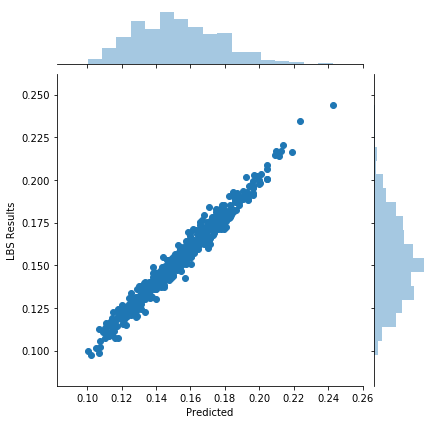}\\
   \caption{Cross-plot of the predicted permeability values for test images vs. ground truth.}
   \label{cross}
 \end{figure}

The test portion of the data, which is not used in the training process, is only used to examine the predictive capabilities and the robustness of the model. 
All the simulations are tested to verify that they have reached steady state conditions, where kinetic energy of the system becomes constant. 
The binary images are used for the LB simulations (Figure \ref{error}a). 
The regions away from the solid-pore interfaces exhibit higher velocity values compared the ones adjacent to the interfaces. Contour plots of x-direction velocity for the developed model (Figure \ref{error}b) and those taken from the LB simulations (Figure \ref{error}c) show similar behavior. The contour plots in Figure \ref{error}c are adjusted based on the minimum and maximum values obtained in contours in Figure \ref{error}b. 
The behavior in Figure \ref{error}b and Figure \ref{error}c are fairly similar, except for the high (hot-colored) values in Figure \ref{error}c. The positions of the circles (zero-velocity valued pixels in Figure \ref{error}b and Figure \ref{error}c) are accurately predicted. 
The error plots (Figure \ref{error}d) exhibit errors in the domain. 
The error for most of the cases are bounded within 20\% errors (Figure \ref{error}e), confirming the plausibility of the approach to replicate costly numerical simulations. 
In this work, a point-by-point comparison of the predicted normalized velocity fields using DDP model and the LB simulations is performed. 
Velocity profiles in a vertical and horizontal cross-section of the simulation domain for one of the test cases are depicted in Figure \ref{profile}.
By inspection of these plots, one can see that DDP model mimics the LB simulation results with negligible errors at points in the solid grain regions (circles). 
Please note that the velocity values at these regions is not zero due to the fact that the minimum value the velocity value in the dataset in transferred to zero, resulting the zero-valued velocities to be transformed into a decimal value (0.07) in the normalized form.
 The DDP predicted velocity values along the vertical cross-section exhibit deviations from the LB simulation results (Figure \ref{profile}b).
 However, at regions close to the interfaces (no-flow boundary), the predictions are close to the LB simulation results.
In horizontal cross-section in Figure \ref{profile}c is selected in a high velocity region. 
 The error in predicted velocity values at inside the grains is negligible. 
 However, the deviations increase as velocity increases.
 In this region, DDP predicts lower values that those calculated using LB simulations. 
The predicted mean of x-direction velocity values for all test cases are shown in Figure \ref{cross}. All the points are projected along the unit-slope line, which shows that predicted values are fairly close to the LB simulation results.
Based on the results, the predicted permeability results for the two-dimensional domains (images) are predicted using the PB-DDP model with high accuracy. 
 
\section{Conclusions}
A data-driven proxy to high-fidelity numerical flow simulations is presented by employing a deep convolutional neural network  based on contracting paths and residual blocks. 
The network consists of only convolutional layers and can take any arbitrary-sized image as input and generate an output of similar size.
 The developed model captures the flow field and permeability for samples at the grid level that had not been used in the development of the model.
Based on the results, the predicted permeability results for the two-dimensional domains (images) are predicted using the PB-DDP model with high accuracy. 
This work aspires to make a transformative leap by combining fluid flow modeling with physics-based machine learning to develop proxy models, which can be used to determine the flow fields at very little additional cost.


\printnomenclature

\bibliography{ref}
\end{document}